\documentclass[10pt, journal, letterpaper, twocolumn]{IEEEtran}
\ifCLASSOPTIONcompsoc
\usepackage{cite}
\else
\usepackage{cite}
\fi

\ifCLASSINFOpdf

\else

\fi
\usepackage{adjustbox}
\usepackage{blkarray, bigstrut}
\usepackage{xparse}
\usepackage[table,xcdraw]{xcolor}
 \usepackage{longtable}
 \usepackage{subcaption}
\usepackage{multirow} 
\usepackage{algorithmicx}
\usepackage{algpseudocode}
\usepackage{algorithm}
\usepackage{rotating}
\usepackage{tabularx}
\usepackage{array,multirow}
\usepackage{color}
\usepackage{cuted}
\usepackage{subcaption}
\usepackage{flushend}
\definecolor{LightCyan}{rgb}{0.88,1,1}
\RequirePackage{graphicx}
\RequirePackage{mathptmx}      
\RequirePackage{flushend}
\usepackage[pdfusetitle, pdfauthor={Michael Shell, My institution}]{hyperref}
\usepackage{mathtools}
\usepackage{amssymb}
\usepackage{amsmath}
\usepackage{calc}
 \usepackage{enumitem}
 \usepackage{graphicx,booktabs}
\usepackage{relsize}
\usepackage{stfloats}

\usepackage{amsthm}

\begin{document}

\hyphenation{op-tical net-works semi-conduc-tor}

 \title{\huge UNBUS: Uncertainty-aware Deep Botnet Detection System in Presence of Perturbed Samples}


\author{Rahim~Taheri
\IEEEcompsocitemizethanks{\protect
\IEEEcompsocthanksitem R. Taheri is with the King’s Communications, Learning and Information Processing (kclip) lab, King’s College London, UK E-mail: rahim.taheri@kcl.ac.uk}}

\markboth{Submitted to Computers and Security Journal}
{Taheri \MakeLowercase{\textit{}}:XXX}


	\maketitle

\begin{abstract}

A rising number of botnet families have been successfully detected using deep learning architectures. While the variety of attacks increases, these architectures should become more robust against attacks. They have been proven to be very sensitive to small but well-constructed perturbations in the input. Botnet detection requires extremely low false-positive rates (FPR), which are commonly attainable in contemporary deep learning. Attackers try to increase the FPRs by making poisoned samples. The majority of recent research has focused on the use of model loss functions to build adversarial examples and robust models. In this paper, two LSTM-based classification algorithms for botnet classification with an accuracy higher than 98\% are presented. Then, an adversarial attack is proposed, which reduces the accuracy to about 30\%. Then, by examining the methods for computing the uncertainty, the defence method is proposed to increase the accuracy to about 70\%. By using the deep ensemble and stochastic weight averaging quantification methods it has been investigated the uncertainty of the accuracy of the proposed methods.
\end{abstract}

\section{Introduction}\label{sez:1}
Nowadays, with the widespread use of computer systems and smartphones, the number of botnets has increased at an unprecedented rate. One of the most popular strategies for detecting botnets are employing machine learning(ML) and deep learning techniques. Numerous studies have been undertaken in recent years in an area known as adversarial deep learning, which seeks to exploit models by taking advantage of available model knowledge and leveraging it to develop adversarial attacks. This architecture is explored from both the attacker and defender perspectives. In this manner, the attacker develops new adversarial techniques while the defender attempts to fight against these attacks using more robust approaches. As a result, the task is to track down and identify vulnerabilities that may be generalized to avoid new attacks and protect computer systems. Despite significant efforts, communities continue to suffer from this problem because the number of botnets grows on a regular basis. This emphasizes the importance of automating botnet detection using ML and deep learning techniques.~\cite{hosseini2022botnet,fazil2021deepsbd}. 

The need for low FPR rates in ML and consequently in deep learning-based techniques has been paramount since the inception of ML detection research~\cite{sadique2022modeling}.

Much of the literature in this domain is focused on cases where the goal was to recognise the set of already known botnets, which makes up a small percent of the total population of botnets. This led to some work that used the training data as part of the final evaluation data~\cite{gandhi2021bond,qiu2022hybrid}.

This approach is not meaningful for determining TPR and FPR due to overfitting and is not tenable due to the large and growing population of botnets with more sophisticated obfuscation techniques. Beyond some works using training data at test time, all of these works evaluate their false-positive rates on the test set, selecting the threshold from the test set that gives them the desired FPR, and then report the associated TPR~\cite{kumar2022machine,yousef2021avoids}. 

This is an understandable but incorrect approach because the threshold is selected explicitly from the test set when our goal is to test the ability of the model to achieve an FPR on unseen data. This is important when we know that misdiagnosis, especially in security systems, sometimes causes irreparable damage to the in-use system. For this reason, the need to use methods that properly reduce the amount of FPR is very important and has attracted a lot of attention in research~\cite{wei2022towards,wang2022manda}.

Prior works have attempted to reduce FPR primarily through feature selection, engineering, or ML processes that they believe will result in a more accurate model or will be biased towards low FPR~\cite{bahcsi2018dimensionality}. Our goal is to better understand how thresholds are chosen to meet TPR and FPR goals and to improve them with uncertainty estimates in the presence of adversarial samples.

In this paper, we use a deep learning approach to detect botnets in two benchmark datasets. In this regard, by selecting the metrics for calculating uncertainty from previous research, the results of botnet detection using the deep learning method have been reviewed and reported. Then an adversarial attack method is proposed that tries to increase the FPR rate by adding perturbation to the data and causing botnets to be misdiagnosed. In the final section, a method for defending against an adversarial attack is proposed, which aims to reduce the FPR rate and increase the accuracy of botnet detection methods. Lastly, the uncertainty of the deep learning botnet detection method is compared in three ways: before the adversarial attack, after the adversarial attack, and after using the botnet defence method.
The contributions of this paper can be summarised as follows:   
        \begin{itemize}[leftmargin=*]
        \item We investigated the prediction uncertainty of the deep learning system in reliably identifying botnets in the presence of poisoning adversarial samples.
        \item Along with the calculation of TPR and FPR, uncertainty is also computed which shows that unlike previous botnet and malware detection tasks that we are aware of them the methods for computing TPR and FPR in the literature are not always suitable.
        \item An adversarial attack has been proposed to determine whether it is effective to use predictive uncertainty to identify adversarial samples.
        \item We have proposed a defence method based on uncertainty metrics against an adversarial attack.
        \end{itemize}
The rest of this paper is organized as follows. First, we will review the related research to our work in section~\ref{relatedWork}. We present the problem definition in section ~\ref{Preliminaries}, and the proposed architecture, attack, and defence  in section ~\ref{ProposedArchitecture}. Finally, simulation setup is discussed in section ~\ref{SimulationSetup}, while performance analysis of the proposed attacks and adapted defence algorithms, is presented in section ~\ref{PerformanceEvaluationResults}. Section ~\ref{conclusion} concludes this paper.

\section{Related Work}\label{relatedWork}
Since the botnet detection methods used in this paper are based on ML, this part first investigates uncertainty in ML and then tries to review some related works in the uncertainty of the botnet detection system before and after adversarial attacks.

Uncertainty induced by a lack of knowledge and inadequate facts required for a perfect predictor is referred to as epistemic uncertainty. It is divided into two categories: approximation uncertainty and model uncertainty. The level of a model's confidence in its prediction is referred to as epistemic uncertainty. The fundamental reason is concern regarding the model's parameters. This form of uncertainty is seen in regions with little training data and incorrectly adjusted model weights. When the model is asked to predict a sample produced from a shifted version of the training data distribution or an out-of-domain sample, there may be a high degree of epistemic uncertainty~\cite{aurigemma2018exploring}.

Aleatoric uncertainty refers to the variation in the outcome of an experiment that is caused by random effects. This kind of uncertainty cannot be eliminated even with an adequate number of training samples. Aleatoric uncertainty, sometimes referred to as statistical uncertainty, is a reflection of unknowns that alter with each repetition of an experiment. Coin flipping is a fundamental example of aleatoric uncertainty: In this kind of experiment, the data generation technique is stochastic.

We divide the related work into two different categories:
\begin{itemize}[leftmargin=*]
\item Uncertainty Estimation 
\item Robustness in Uncertainty Estimation 
\end{itemize}

\subsection{Uncertainty Estimation}\label{UncertaintyEstimation}  
Various studies have considered methods for estimating uncertainty~\cite{renaud2021uncertainty}. In~\cite{van2020uncertainty}, the authors offer a technique for training a deterministic deep model with a single forward pass that can discover and reject data points that are outside of the distribution at testing time. Their method, called "deterministic uncertainty quantification," is based on RBF network ideas and scales by employing a unique loss function and centroid updating strategy to calculate uncertainty. 

The authors in ~\cite{wortwein2020simple} look at methods for predicting absolute error to encompass both epistemic and aleatoric uncertainties. They consider uncertainty to be a weighted average of known absolute errors from similar reference data. In~\cite{mobiny2021dropconnect}, by imposing a Bernoulli distribution on the model weights, a theoretical framework for Bayesian inference for Deep Neural Networks(DNNs) is developed. Monte Carlo DropConnect (MC-DropConnect) is a method that they used and allows for describing model uncertainty with minimal changes to the overall model structure or computational cost.

 The authors of ~\cite{tsiligkaridis2021information} propose a new  Information aware Dirichlet network. These networks by minimizing a bound on the expected max norm of the prediction error and penalizing information associated with incorrect predictions, learn an explicit Dirichlet prior distribution on predictive distributions. Based on this work, for guaranteeing trust and safety, precise quantification of uncertainty in AI system predictions is crucial.

One application of uncertainty is presented in~\cite{qiao2020uncertain}. Based on this work out-of-domain generalization from a single source is the worst-case scenario in generalization. The goal of this work is to learn a robust model from a single source that can generalize across a large number of unknown distributions. The main aim is to increase source capacity in both the input and label spaces, guided by the uncertainty assessment. In this study two points can be seen: former, access generalization uncertainty from a single source and later, using it to drive both input and label augmentation for robust generalization.

According to the author's opinions in ~\cite{krishnan2019efficient}, the specification of priors and approximate posterior distributions for neural network weight is required for stochastic variational inference for Bayesian DNN. It's challenging to define meaningful weight priors, especially when expanding variational inference to deeper architectures with high dimensional weight space.
The authors' contribution is a simple but effective method for initializing weight priors in DNNs to enable scalable variational inference. They propose to establish meaningful priors for DNNs based on the deterministic weights of pre-trained DNN models, determined from maximum likelihood estimations.

\subsection{Robustness in Uncertainty Estimation} \label{RobustnessUncertaintyEstimation} Different ML algorithms are exposed to adversarial samples, and there are different ways to create and defend against adversarial attacks. Uncertainty measurement is a potential technique for finding adversarial cases, which are prepared inputs for which the model predicts an incorrect class with a high degree of certainty. There are, however, many different types of uncertainty metrics, such as predictive entropy and mutual information, each of which represents a different type of uncertainty.Researchers in some studies have used uncertainty estimates for attack and defence against ML models~\cite{anthi2021hardening}.

In ~\cite{zhou2021amortized} authors proposed a method for uncertainty estimates, calibration, and out-of-distribution robustness using deep networks. Their solution is based on the conditional normalized maximum likelihood coding scheme, which is computationally challenging to evaluate. Using approximate Bayesian inference techniques, they proposed building a tractable approximation to the conditional normalized maximum likelihood distribution. The authors in this study investigate how to obtain accurate uncertainty estimates when the distribution shifts, with the goal of developing models that can accurately report their uncertainty even when given unexpected inputs.

To improve the robustness of neural networks, authors in  ~\cite{zhang2021robust} propose spectral expectation bound regularization. According to this theoretical study, training with this method improves robustness to adversarial samples. They also show that training using spectral expectation bound regularization reduces the model's epistemic uncertainty and, as a result, makes the model more confident in its predictions. According to the theorem that is presented in this work the proposed spectral expectation bound regularization efficiently minimizes the epistemic uncertainty on the output of the Bayesian neural network. On Bayesian neural networks trained with spectral expectation bound regularization and without it, they assess the aleatoric uncertainty and epistemic uncertainty to confirm the resilience of the models. In this research, the authors investigated Bayesian learning for generative adversarial networks, which addressed the problem of mode collapse by regularizing trained models and compensating for weight uncertainty. A new Bayesian GAN is proposed and implemented to learn a regularized model from a variety of data, with the strong modes flattened by marginalization and model collapse and gradient vanishing issues tackled. This work introduces a variational GAN that uses variational Bayesian inference to estimate the encoder, generator, and discriminator at the same time. In fact, a variational GAN explores the posterior across parameters and generates realistic synthesized samples by maximizing the variational lower bound of log-likelihood.

The authors in ~\cite{qiao2018toward} offer a method for adversarial samples using a large margin cosine estimate. Their results may be regarded as a novel measurement of model uncertainty estimation and are available to detect adversarial samples by training a ML algorithm by iteratively calculating the large-margin cosine uncertainty estimates between the model predictions. In this research, it is confirmed that this measurement can better distinguish adversarial perturbations by considering the manner in which adversarial samples are generated. For example, DNNs face difficulty when learning in uncertain, noisy, or adversarial environments.

A new theoretically supported and efficient approach for robust learning based on Bayesian estimation and variational inference, is presented in~\cite{carannante2020robust}. The authors use an ensemble density propagation approach to handle the problem of density propagation through the layers of a DNN. This method allows propagating moments of a variational probability distribution across the layers of a DNN, allowing one to estimate the predictive distribution's mean and covariance at the model's output.

A framework for jointly training deep generative and discriminative models is proposed by the authors in ~\cite{gordon2020combining}, which takes advantage of both approaches. They claimed that no Bayesian approach to semi-supervised learning using deep generative models has been developed that allows models to learn from labelled and unlabeled data as well as account for uncertainty in predicted distributions.

The authors of ~\cite{feinman2017detecting} use Bayesian uncertainty estimates from dropout neural networks and density estimation in the subspace of deep features learned by the model to examine model confidence on adversarial samples. As a result, the attack algorithm is blind to the implicit adversarial detection mechanism. They showed that by utilising two novel features, i.e., kernel density estimates in the last hidden layer's subspace and Bayesian neural network uncertainty estimates, generated adversarial samples to fool DNNs can be recognised effectively. These two features deal with different conditions and can be combined to make a successful defence against adversarial samples.

There are two gaps in previous works that this paper attempts to address. First, the use of uncertainty estimation methods has not been used to propose adversarial attack and defence algorithms, and the accuracy of the proposed methods has not been investigated considering uncertainty. Second, in the field of botnets and malware, no one has looked at how likely the models are.

\section{Preliminaries}\label{Preliminaries}
Predictive models are used in many cases of ML research to develop decisions that are not reliable and are expected to have low quality. As stated in related work, a lack of uncertainty consideration in ML methods can result in unreliable results. As a consequence, several researchers have attempted to quantify uncertainty and have proposed models for high-uncertainty scenarios in order to increase reliability. On the other hand, detecting botnets is also an important issue in computer security, and it is critical to highlight the uncertainty in the proposed approaches for this purpose, especially when the detection system is under attack.

In this paper, different methods of uncertainty quantification are considered. First, an adversarial attack on the botnet detection system is proposed, and uncertainty quantification methods a defence  algorithm is presented. In this detection system, in addition to comparing the results with the usual ML metrics, uncertainty measures are also used for comparison.

\subsection{General Metrics for Uncertainty Quantifying}\label{UncertaintyMetrics}
The following general metrics have been proposed in different works to quantify uncertainty.

\begin{itemize}[leftmargin=*]
\item{\textbf{Entropy value:}}
To measure neural network uncertainty for class predictions, the entropy of output classes is calculated as eq~\ref{eq:Entropy}. $N$ is number of classes:
\begin{equation}
\label{eq:Entropy}\small
H\left(p\left(y \mid x, w\right)\right)=-\sum_{k=1}^{N} p\left(y_{i} \mid x, w\right) \cdot \log p\left(y_{i} \mid x, w\right)
\end{equation}
In this equation, $p\left(y_{i} \mid x, w\right)$ indicates the probability of label $y_i$ while sample($x$) and weight$w$ vectors are given. Entropy is calculated at each time step and its maximum value is used as a measure of uncertainty. This value can be used to compute the uncertainty of all types of neural networks, including deep neural networks.

\item{\textbf{Mutual Information Value:}}
 The mutual information value is computed based on the entropy value. In this way, $M$ different predictions are calculated, and then the mutual information value is obtained according to the equation~\ref{eq:MutualInformation}. In this case, the calculation is based on $M$ different predictions, each with its own weights $w_{i}$.
\begin{equation}
\label{eq:MutualInformation}\small
\mathrm{MI}=\mathcal{H}[p(y \mid x)]-\frac{1}{M} \sum_{i=1}^{M} \mathcal{H}\left[p\left(y \mid x, w_{i}\right)\right]
\end{equation}
Similar to the entropy method, a higher MI value indicates higher uncertainty.

\item{\textbf{Variance Value:}}
This measure is based on the variance of different predictions and computed as the equation~\ref{eq:Variance}. $p\left(y_{i} \mid x, w_{i}\right)$ indicates the probability of label $y$ while sample($x$) and weight$w_{i}$ vectors are given. $M$ is number of predictions.
\begin{equation}
\label{eq:Variance}\small
VV=\frac{1}{M} \sum_{i=1}^{M} p\left(y \mid x, w_{i}\right)^{2}-p(y \mid x)^{2}
\end{equation}

\item{\textbf{Kullback-Leibler Divergence:}}
The Kullback-Leibler divergence is another measure which its idea is to observe changes in distributions. The average of this divergence called AKLD and is calculated by equation~\ref{eq:KullbackLeibler}:
\begin{equation}
\label{eq:KullbackLeibler}\small
AKLD=\frac{1}{M-1} \sum_{i=1}^{M-1} p\left(y \mid x, w_{i}\right) \log \frac{p\left(y \mid x, w_{i}\right)}{p\left(y \mid x, w_{i+1}\right)}
\end{equation}
\end{itemize}

\subsection{Uncertainty Estimation in Deep Learning}
In this subsection, we will quickly outline the two approaches for estimating uncertainty that we employed in this work.

Suppose the $D$ dataset as follows is assumed to contain $d$ botnet instances which all of them were sampled from the same data generator distribution.

\begin{equation}
\label{eq:Data}\small
D:=\left\{\left(x_{i}, y_{i}\right)\right\}_{i \in d}  
\end{equation}

In a DNN, a common way to predict the output value of $y$ is to find $Y$ with the highest probability. i.e:\\
\begin{equation}
\label{eq:argmaxmodel}\small
\operatorname{argmax}_{y} p\left(Y=y \mid x, \theta\right).
\end{equation}
In this regard, $p\left(Y=y \mid x, \theta\right)$ is the model obtained from DNN training on $D$ with $\theta$ as DNN parameters, which is the conditional probability distribution of output $y$ while input $x$ is given.

\begin{itemize}[leftmargin=*]
\item{\textbf{Deep Ensembles:}} 
Ensembles have been widely explored in ML and  are the simplest ways for estimating uncertainty in neural networks.  In practise, it is recognised as a technique for enhancing the performance of weak classifiers, including neural networks and other models such as random forests.
One of the trivial ensemble approaches proposes training a neural network $S$ times with various random initializations, resulting in $S$ distinct models with parameter sets $\theta_{1}, \ldots, \theta_{S}$. For the final result, the average of all the projections will compute by equation~\ref{eq:ensemble}:
\begin{equation}
\label{eq:ensemble}\small
p\left(y \mid x\right)=\frac{1}{S} \sum_{s=1}^{S} p\left(y \mid x, \theta_{s}\right).
\end{equation}

By computing $p\left(y \mid x\right)$, we can consider uncertainty by the variance between network predictions as equation~\ref{eq:uncertaintyvariance}.
\begin{equation}
\label{eq:uncertaintyvariance}\small
\sigma^{2}=\frac{1}{S} \sum_{s=1}^{S} p\left(y \mid x, \theta_{s}\right)^{2}-p\left(y \mid x\right)^{2}
\end{equation}

\item{\textbf{Stochastic Weight Averaging:}}
Another approximate Bayesian method called SWAG uses stochastic gradient descent. This method is inspired by Stochastic Weight Averaging (SWA) and starts with a pre-trained solution. Then averaging on network parameters in the stochastic gradient descent pathway improves generalization. Where the parameters are used with a normal distribution, the average of each parameter during training such as SWA is calculated by equation~\ref{eq:SWAG}.
\begin{equation}
\label{eq:SWAG}\small
\theta_{\mathrm{SWA}}=\frac{1}{T} \sum_{i=1}^{T} \theta_{i}
\end{equation}
In this equation $T$ is the number of SWA epochs. SWAG computes the covariance of the parameters by averaging the second uncentered moment of each weight by equation~\ref{eq:seconduncentered}:
\begin{equation}
\label{eq:seconduncentered}\small
\overline{\theta^{2}}=\frac{1}{T} \sum_{i=1}^{T} \theta_{i}^{2}
\end{equation}

Following the final training stage, equation~\ref{eq:covariancematrix} combines $\overline{\theta^{2}}$ and $\theta_{\text {SWA }}$ to formulate a covariance matrix over the parameters.
\begin{equation}
\label{eq:covariancematrix}\small
\Sigma_{\text {diag }}=\operatorname{diag}\left(\overline{\theta^{2}}-\theta_{\mathrm{SWA}}^{2}\right) .
\end{equation}
So, the resulting parameter distribution $\mathcal{N}\left(\theta_{\mathrm{SWA}}, \Sigma_{\text {diag}}\right)$ is considered as $p(\theta \mid D)$.
\end{itemize}

\section{Proposed adversarial attack and Defence}\label{ProposedArchitecture}
In this section, the proposed adversarial attack and defence algorithms are presented.
\subsection{Weight-based adversarial attack}
The purpose of this type of poisoning attack is to change the weights of the model in such a way that it is trained by the samples that cause the higher loss. In this type of attack, the adversary selects samples of training data that cause a higher loss and considers them to generate poisoning data. The "weight update" step in network training is divided into two parts. For generating each poisoning sample, in addition to applying changes due to its weight, total weight changes are also involved, while for other training data, only weight changes in samples are used. As a result, the model produced will tend to sample with higher losses. This higher loss in detecting botnets of test data will lead to a higher rate of incorrect classification of samples.

In Algorithm~\ref{alg:attack}, we consider $\epsilon$ as the rate of the training data that we want to poison. In lines 4 to 7 of the algorithm, we select the samples with the highest loss from the training data and collect them in $D_{Poison}$. Then, in the weights update step, by considering $\lambda$ as the rate we update the weights in lines 8 to 10 based on $D_{train}$ and in lines 11 to 13 based on $D_{Poison}$. The output of this algorithm will be the weights that we will use to train and generate a poison model.

\begin{algorithm}[t]
	\footnotesize
\caption{\small WBA: Weight-based adversarial attack}
\label{alg:attack}
\textbf{Input:} $D_{train}$,$\epsilon$,$N$ \\
\textbf{Output:} $w$
 
\begin{algorithmic}[1]
\color{black}
\footnotesize
\State{Initialize $w \in R$}
\State{$D_{Poison} \leftarrow \varnothing$}
\State{$P\leftarrow \epsilon N$}
\For{each $(x,y) \in D_{train}$}
\State{$(x_p,y_p)\leftarrow$ \text{Pop}  $\underset{(x,y) \in D_{\text {train }}}{\operatorname{argmax}  \ell}(w,x,y)$}
\State{$D_{Poison}=D_{Poison} \cup (x_p,y_p)$}
\EndFor
\For{each $(x,y)$ in $D_{train}$}
\State{${w} \leftarrow {w}-\lambda \nabla \mathcal{L}(x, y,w)$} 
\EndFor
\For{each $(x,y)$ in $D_{Poison}$}
\State{${w} \leftarrow {w}-\lambda \left[\nabla \mathcal{L}(x, y,w)+ \nabla \mathcal{L}(w)\right]$} 
\EndFor
\State{\textbf{return $w$}} 
\end{algorithmic}
\end{algorithm}

\subsection{Uncertainty metric based defence}
In this section, a defence method based on uncertainty quantification is proposed, which uses uncertainty calculation measures. In order to identify adversarial samples, we calculate the uncertainty for each sample of the attacked dataset with each of the measures introduced in section~\ref{UncertaintyMetrics} and select $(1-\epsilon)\%$  of samples that have lower values based on these measures. In fact, we select those with lower uncertainties and train the model with these examples. The main idea of this method is that adversarial samples are often part of the $\epsilon\%$ of data with high uncertainty. This method is presented in Algorithm~\ref{alg:Defence}.

\begin{algorithm}[t]
	\footnotesize
\caption{\small UMD: Uncertainty metric based defence}
\label{alg:Defence}
\textbf{Input:} $D$,$N$,$\epsilon$ \\
\textbf{Output:} $D_{corrected}$

\begin{algorithmic}[1]
\color{black}
\footnotesize
\State{$D_{corrected} \leftarrow \varnothing$}
\State{$P\leftarrow \epsilon N$}
\For{each $c \in P$}
\State{A= Compute Entropy value by equation~\ref{eq:Entropy}}
\State{MI= Compute Mutual Information Value by equation~\ref{eq:MutualInformation}}
\State{VV= Compute Variance Value by equation~\ref{eq:Variance}}
\State{AKLD= Compute Kullback-Leibler Divergence by equation~\ref{eq:KullbackLeibler}}
\State{$(x_c,y_c)\leftarrow$ \text{Pop}  $\underset{(x,y) \in D}{\operatorname{argmax}  \ell}(A,MI,VV,AKLD)$}
\State{$D_{corrected}=D_{corrected} \cup (x_c,y_c)$}
\EndFor
\State{\textbf{return $D_{corrected}$}} 
\end{algorithmic}
\end{algorithm}

\subsubsection{Time Complexity of proposed Attacks and Defence }
The time complexity of the proposed adversarial attack method depends on the three separate 'for' loops and the operations within them. Given that these three loops are separate, each of them has $\mathcal{O}(n)$ complexity, and by considering the derivative operation of all samples with $\mathcal{O}(n)$ complexity, the time complexity of the attack method is $\mathcal{O}(n^2)$.
The time complexity of the proposed defence method depends on the 'for' loop and thus the number of adversarial samples. Because this loop is running as a fraction of the number of adversarial samples, although the running time is greatly reduced, the time complexity of the 'for' loop is still$\mathcal{O}(n)$. If we consider the time complexity of calculating the equations~\ref{eq:Entropy} to~\ref{eq:KullbackLeibler}, which in the worst case is $\mathcal{O}(n)$, the time complexity of the defence method is $\mathcal{O}(n^2)$.

\section{Simulation Setup}\label{SimulationSetup}
In the following subsections, we present the classification algorithms used for train models, botnet datasets, and ML test metrics used in this paper.
\subsection{Classification Models}
In this section, we introduce the classification architectures used for botnet detection, which are used for the evaluation of uncertainty and the effect of adversarial samples on it. According to the methods that were used in the botnet literature, both proposed architectures are based on LSTM.

\begin{itemize}[leftmargin=*]
\item{\textbf{LSTM Based Classifier:}}
In designing this classifier, 10 LSTM cells have been used, which are connected sequentially. Using the embedding layer and the dropout layers before and after the LSTM layer was just one of many possible choices, and no attempt was made to make the best choice. In the dropout layers, the rate is 0.5, and finally in the fully connected layer, the standard softmax function is used to classify the samples into two classes: benign and botnet. This classifier is presented in Figure~\ref{fig:LSTM}.
\begin{figure}[!htb]
\centering 
\includegraphics[width=0.95\columnwidth]{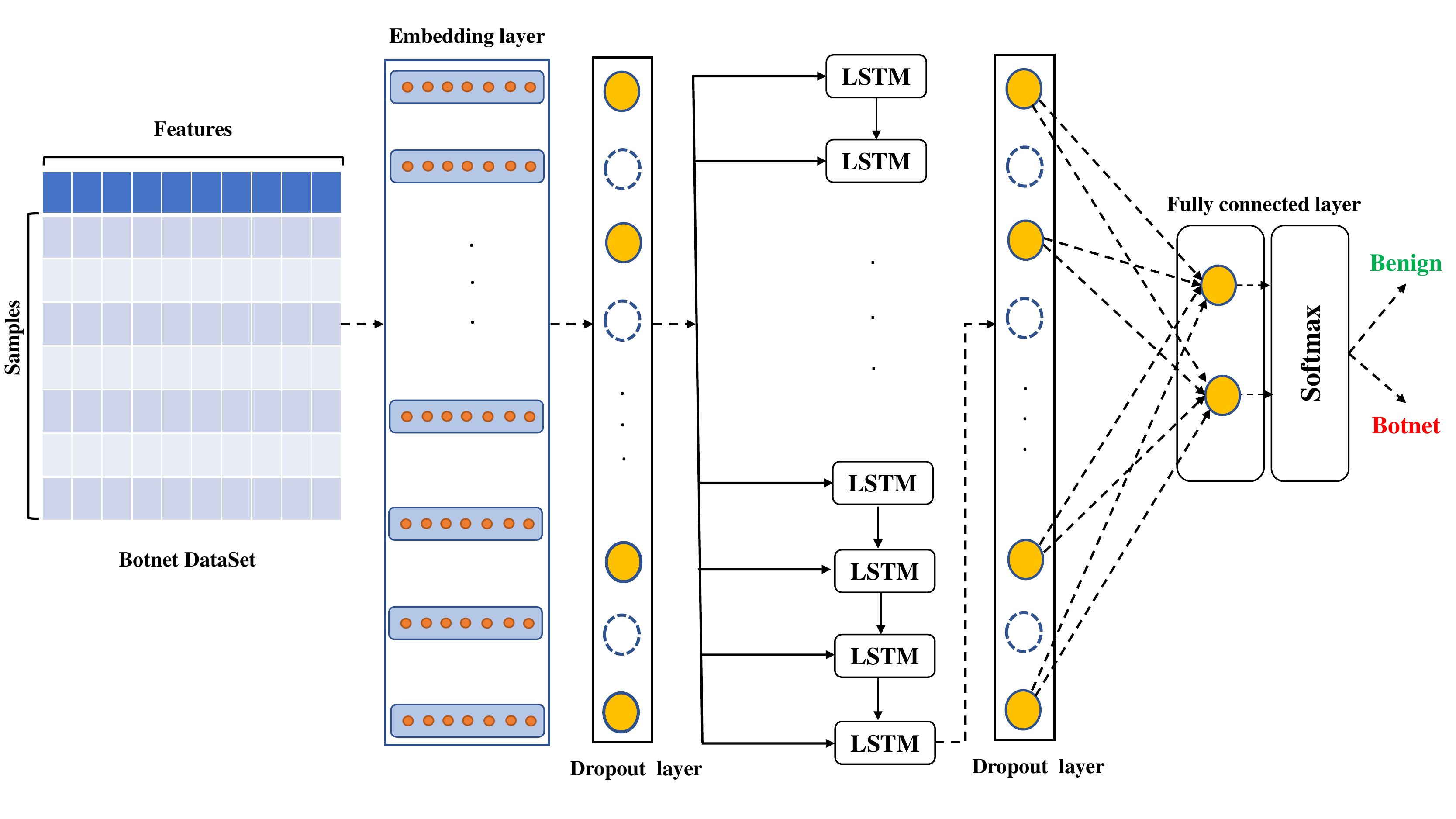}
\caption{\small LSTM Based Classifier Architecture}
\label{fig:LSTM}
\end{figure}

\item{\textbf{CNN and LSTM Based Classifier:}}
The classifier is designed similarly to the LSTM classifier, and just a convolution1D layer is added before the LSTM cell layer. Convolution1D uses ReLU as the activation function and a MaxPooling1D with size 2 which is added sequentially. Other layers and hyperparameters are similar to the LSTM Classifier. This classifier is presented in Figure~\ref{fig:CNN-LSTM}.
\begin{figure}[!htb]
\centering 
\includegraphics[width=0.95\columnwidth]{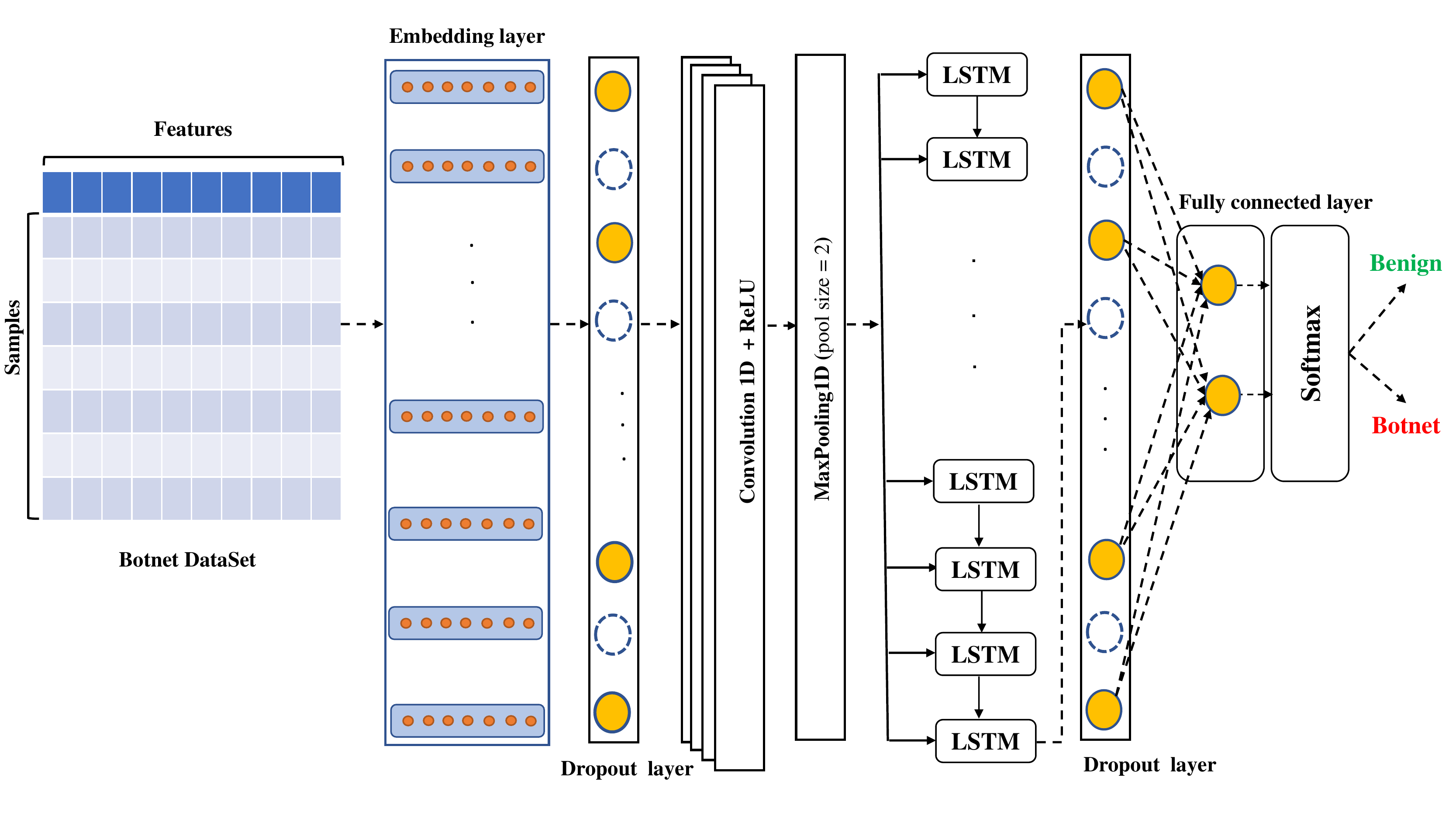}
\caption{\small CNN and LSTM Based Classifier Architecture}
\label{fig:CNN-LSTM}
\end{figure}
\end{itemize}

\subsection{Datasets}\label{Datasets}
Two datasets with the following specifications have been used to test the proposed methods.
\begin{itemize}[leftmargin=*]
\item{\textbf{N-BaIoT Dataset:}} 
In this dataset~\cite{BaIoT}, botnets are generated and managed in several stages, including propagation, bot infection, communication with the server, and other malicious activities. This dataset is a real dataset collected from IoT network traffic and includes labeled malicious and benign samples. This is a collection of real traffic data from 9 IoT devices infected with Mirai and Bashlite. This dataset is collected from networks of large organizations that are expected to see an increasing number of IoT devices. In this research, 108623 samples of this data set, each of which has 115 features, have been used, and 70\% of this data has been used as a training dataset. In this data set, there is a balance between benign and botnet samples, i.e., half of the data is labeled by each of these two types.

\item{\textbf{IOT-23 Dataset:}}
This dataset~\cite{IOT-23} contains 20 types of botnet samples collected from IoT devices plus three types for benign IoT device traffic, first released in January 2020. The purpose of this dataset is to provide a wide range of real-world IoT malicious infections that are labelled and used to develop ML algorithms. This paper uses data from this data set, which has 311913 samples and each sample has 10 features. This dataset is labelled and contains labels to describe the relationship between flows related to potential malicious activity. This labeling was created by analyzing and capturing botnets in the Stratosphere lab and the number of benign and botnet data is balanced. Similar to the N-BaIoT dataset, 70\% of the data in this dataset is used as training data. 

\end{itemize}
\textbf{Implementation Environment:}
The tests in this paper were performed on an eight-core Intel Core i7 computer clocked at 4 GHz. This computer uses 64-bit Win10 and 16 GB of RAM. All implementations are done in Python 3.

\section{Performance Evaluation and Results}\label{PerformanceEvaluationResults}
For performance evaluation of the proposed attack and defence methods in this paper, the ML metrics mentioned in Table~\ref{PerformanceEvaluation} have been used and all of the equations mentioned in this table are based on the confusion matrix and includes:  
\begin{itemize}[leftmargin=*]
\item{\textbf{TP:} The number of examples successfully classified as belonging to the class.}
\item{\textbf{TN:} It indicates the number of negative examples which are correctly classified.}
\item{\textbf{FP:} Samples that were misclassified as belonging to the class.}
\item{\textbf{FN:} The number of positive examples classified as negative.}
\end{itemize}
\begin{table}[!htpb]
\centering
\caption{\small   {ML metrics used in this paper for performance evaluation }\vspace{-10px}}
\label{PerformanceEvaluation}
\scriptsize{
\setlength\tabcolsep{6pt} 
\begin{tabular}{|c|
>{\columncolor[HTML]{fcfcf4}}l|}
\hline
\rowcolor{LightCyan}
\textbf{Metric}  
&\multicolumn{1}{|c|}{\textbf{Equation}}\\ \hline
\cellcolor[HTML]{E8E8AB}\textbf{Accuracy}& $\frac{TP+TN}{TP+TN+FP+FN}$  \\\hline
\cellcolor[HTML]{E8E8AB}\textbf{Precision}& $\frac{TP}{TP+FP}$\\\hline
\cellcolor[HTML]{E8E8AB}\textbf{Recall}& $\frac{TP}{TP+FN}$ \\\hline
\cellcolor[HTML]{E8E8AB}\textbf{FPR}&$\frac{\mathrm{FP}}{\mathrm{FP}+\mathrm{TN}}$ \\\hline
\cellcolor[HTML]{E8E8AB}\textbf{TPR}& $\frac{\mathrm{TP}}{\mathrm{TP}+\mathrm{FN}}$ \\\hline
\cellcolor[HTML]{E8E8AB}\textbf{F1-Score}& $2 \cdot \frac{\text { Precision } \cdot \text { Recall }}{\text { Precision }+\text { Recall }}$ \\\hline
\end{tabular}}
\end{table}

\subsection{Comparing Based on ML Metrics}
\begin{figure}[!htb]
\centering 
\includegraphics[width=0.95\columnwidth]{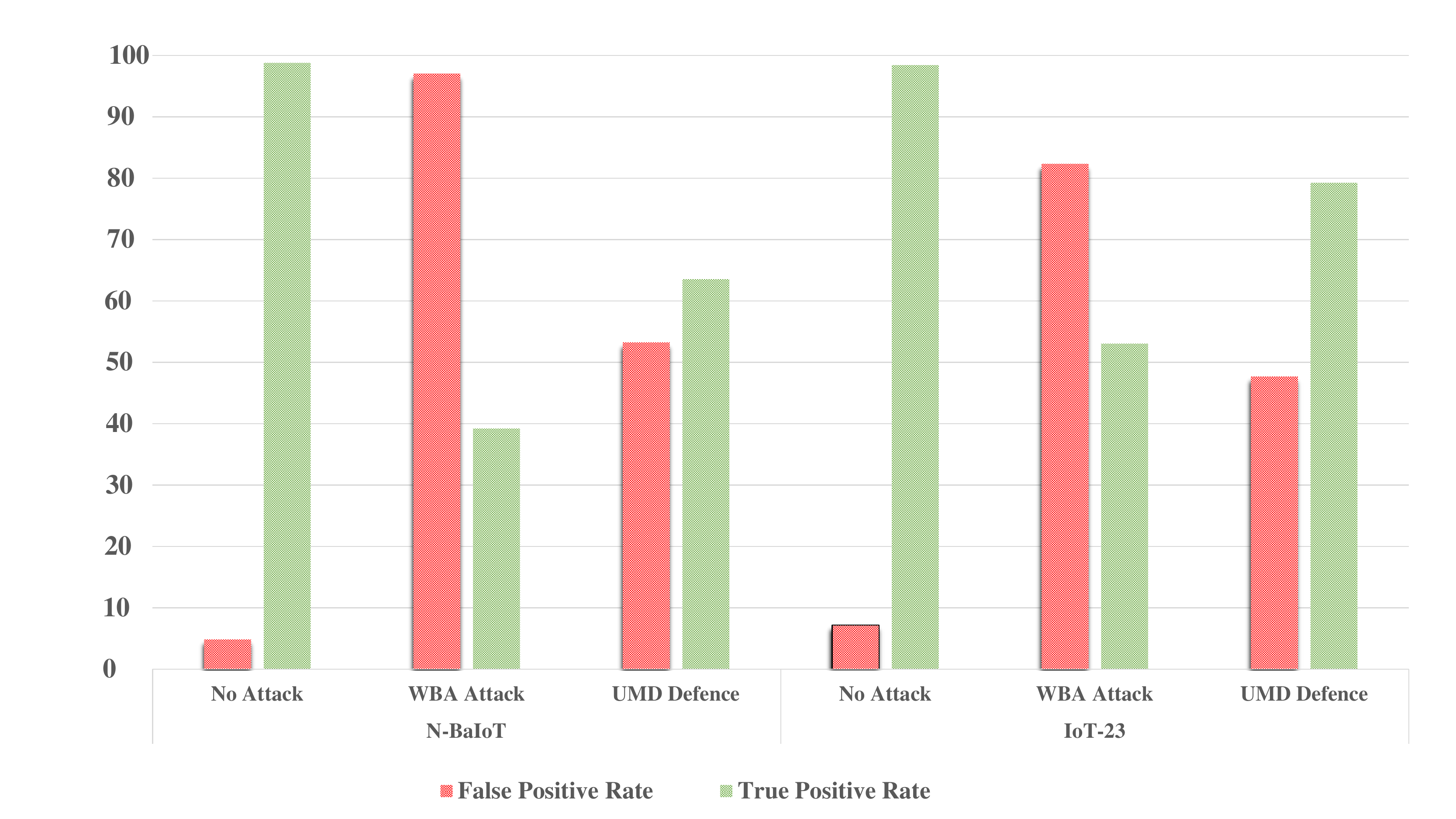}
\caption{\small LSTM Based Classifier\vspace{-1px}}
\label{fig:LSTMFalsePositiveTruePositive}
\end{figure}

Table~\ref{tab:LSTMClassifierMLMetrics} presents the results for the classification of botnets in the N-BaIoT and IoT-23 datasets using the LSTM Classifier proposed in this paper. As expected, the "No Attack" line indicates a state in which no attack has taken place, and only the LSTM Classifier is used to make a model for the classification of data, which has high accuracy values (98.56\% for the N-BaIoT data set and 98.32\% for the IoT-23 dataset). The WBA Attack in the two datasets shows the state in which the adversarial attack proposed in this paper was performed (the results of this table are for 10\% change in training data). In the event that the attack occurs, accuracy, precision, recall, and f1-Score metrics decrease, and FPR increases, as shown in Figure~\ref{fig:LSTMFalsePositiveTruePositive}. Conversely, using UMD defence, it is observed that in both datasets, the accuracy of the other metrics mentioned above increases, and the FPR value decreases according to Figure~\ref{fig:LSTMFalsePositiveTruePositive}.

\begin{table}[!htpb]
\centering
\caption{ \small LSTM Classifier\vspace{-10px} }
\label{tab:LSTMClassifierMLMetrics}
\scriptsize{
\setlength\tabcolsep{0.75pt} 
\begin{tabular}{|c|c||c||c||c|}
\hline
\rowcolor{LightCyan}
\multicolumn{5}{|c|}{\textbf{N-BaIoT Dataset}}\\\hline\hline
\rowcolor[HTML]{E8E8AB}
\multirow{-1}{*}{\textbf{Scheme}} &\multicolumn{1}{|c||}{\textbf{Accuracy}}&\multicolumn{1}{|c||}{\textbf{Precision}}&\multicolumn{1}{|c|}{\textbf{Recall}}&\multicolumn{1}{|c|}{\textbf{F1-Score}}\\\cline{2-5}
\textbf{No Attack}&98.56&99.69&98.78&99.23\\\hline
\textbf{WBA Attack}&28.46&48.93&39.22&43.54\\\hline
\textbf{UMD Defence}&62.54&94.96&63.54&76.14\\\hline\hline

\rowcolor{LightCyan}
\multicolumn{5}{|c|}{\textbf{IoT-23 Dataset}}\\\hline\hline
\rowcolor[HTML]{E8E8AB}
\multirow{-1}{*}{\textbf{Scheme}} &\multicolumn{1}{|c||}{\textbf{Accuracy}}&\multicolumn{1}{|c||}{\textbf{Precision}}&\multicolumn{1}{|c|}{\textbf{Recall}}&\multicolumn{1}{|c|}{\textbf{F1-Score}}\\\cline{2-5}
\textbf{No Attack}&98.32&99.86&98.42&99.14\\\hline
\textbf{WBA Attack}&31.03&28.12&53.05&36.76\\\hline
\textbf{UMD Defence}&73.56&86.12&79.25&82.55\\\hline

\end{tabular}}
\end{table}

\begin{figure}[!htb]
\centering 
\includegraphics[width=0.95\columnwidth]{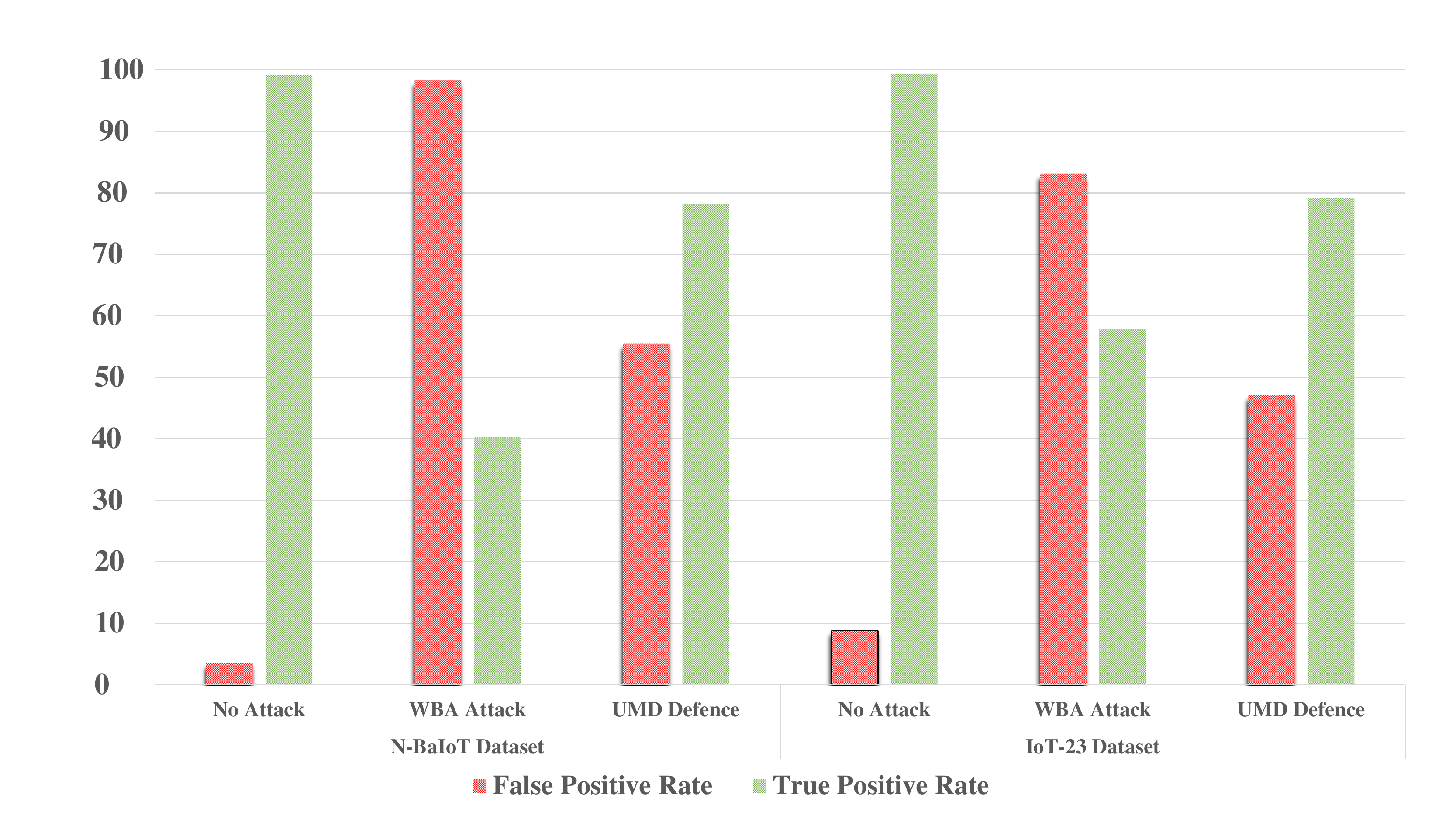}
\caption{\small CNN-LSTM Based Classifier\vspace{-1px}}
\label{fig:CNN-LSTMFalsePositiveTruePositive}
\end{figure}

Table~\ref{tab:CNN-LSTMClassifierMLMetrics} presents the results for the classification of botnets in the N-BaIoT and IoT-23 datasets using the CNN-LSTM Classifier proposed in this paper. Similar to the previous method, the "No Attack" indicates a state in which no attack has taken place, and just a trained model from the CNN-LSTM Classifier is used for data classification, which has high accuracy values (99.02\% for the N-BaIoT data set and 99.21\% for the IoT-23 dataset). Applying WBA Attack to this classification method and using it for botnet classification in the two datasets shows that accuracy, precision, recall, and f1-Score metrics decrease (similarly, the results of this table are for 10\% change in training data), and FPR increases as shown in Figure~\ref{fig:CNN-LSTMFalsePositiveTruePositive}. Similar to the LSTM classifier, by using UMD defence, it is observed that in both datasets, the accuracy of the other metrics mentioned above increases, and the FPR value decreases according to Figure~\ref{fig:CNN-LSTMFalsePositiveTruePositive}.

\begin{table}[!htpb]
\centering
\caption{ \small CNN-LSTM Classifier\vspace{-10px}}
\label{tab:CNN-LSTMClassifierMLMetrics}
\scriptsize{
\setlength\tabcolsep{0.75pt} 
\begin{tabular}{|c|c||c||c||c|}
\hline
\rowcolor{LightCyan}
\multicolumn{5}{|c|}{\textbf{N-BaIoT Dataset}}\\\hline\hline
\rowcolor[HTML]{E8E8AB}
\multirow{-1}{*}{\textbf{Scheme}} &\multicolumn{1}{|c||}{\textbf{Accuracy}}&\multicolumn{1}{|c||}{\textbf{Precision}}&\multicolumn{1}{|c|}{\textbf{Recall}}&\multicolumn{1}{|c|}{\textbf{F1-Score}}\\\cline{2-5}
\textbf{No Attack}&99.02&99.78&99.18&99.48\\\hline
\textbf{WBA Attack}&30.76&55.64&40.25&46.71\\\hline
\textbf{UMD Defence}&71.94&86.01&78.23&81.94\\\hline\hline
\rowcolor{LightCyan}
\multicolumn{5}{|c|}{\textbf{IoT-23 Dataset}}\\\hline\hline
\rowcolor[HTML]{E8E8AB}
\multirow{-1}{*}{\textbf{Scheme}} &\multicolumn{1}{|c||}{\textbf{Accuracy}}&\multicolumn{1}{|c||}{\textbf{Precision}}&\multicolumn{1}{|c|}{\textbf{Recall}}&\multicolumn{1}{|c|}{\textbf{F1-Score}}\\\cline{2-5}
\textbf{No Attack}&99.21&99.34&99.36&99.60\\\hline
\textbf{WBA Attack}&32.29&29.57&57.79&39.13\\\hline
\textbf{UMD Defence}&76.4&92.63&79.13&85.35\\\hline

\end{tabular}}
\end{table}

One of the important comparisons made in hostile ML is the comparison of FPR and TPR. In Figures~\ref{fig:LSTMFalsePositiveTruePositive} and~\ref{fig:CNN-LSTMFalsePositiveTruePositive}, the values of these two metrics are compared for the two classification methods (LSTM and CNN-LSTM) and the two datasets (N-BaIoT and IoT-23). In both figures, it can be seen that before the attack the FPR value is low and the TPR value is high, and after the attack, the FPR increase and the TPR decrease. In both data sets and for both classification methods, it is observed that the defence method leads to a decrease in FPR and an increase in TPR, but can never approach the values before the attack.

A comparison of the results presented in Tables~\ref{tab:LSTMClassifierMLMetrics} and~\ref{tab:CNN-LSTMClassifierMLMetrics} and Figures~\ref{fig:LSTMFalsePositiveTruePositive} and ~\ref{fig:CNN-LSTMFalsePositiveTruePositive} show that the proposed attack method has succeeded in reducing the classification accuracy to about 30\% and the proposed defence method which is based on uncertainty metrics has been able to increase model accuracy to above 70\% in most cases. Between the two classification algorithms used, CNN-LSTM has better results, and according to the results, the used models seem to work better on the IoT-23 dataset.

\subsection{Comparing Based on Different amount of perturbation}
\begin{figure}[!htb]
\centering 
\includegraphics[width=0.95\columnwidth]{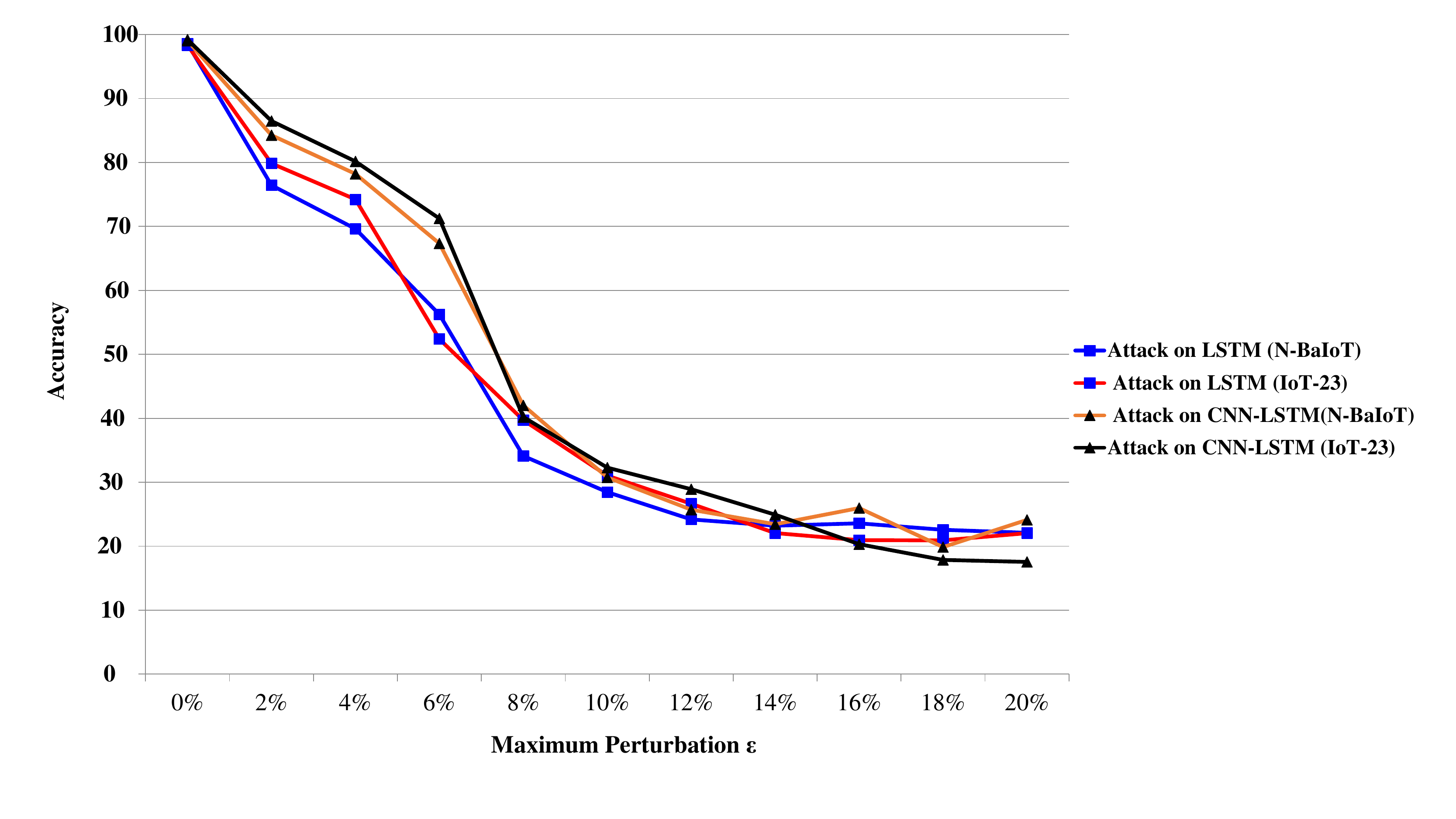}
\caption{\small Maximum Perturbation Vs Accuracy}
\label{fig:MaximumPerturbationAccuracy}
\end{figure}

In Figure~\ref{fig:MaximumPerturbationAccuracy}, the accuracy of the proposed attack method in two data sets and by using two classifiers with different amounts of adversarial samples are shown. As can be seen, with the increasing rate of adversarial samples, the accuracy has decreased drastically, and when just 10\% of the data are adversarial samples, the accuracy has reached about 20\% (although in different classifiers and datasets, this number is different, but the slope of decreasing accuracy is the same). In increasing the rate of adversarial samples from 10\% to 20\%, the slope has decreased.

\subsection{Comparing Based on Uncertainty Quantifying Method}
The results of using two uncertainty quantifying methods, namely deep ensemble and stochastic weight averaging for both proposed classifiers and calculating accuracy based on them are presented in Tables~\ref{table:UncertaintyLSTMClassifier} and~\ref{table:UncertaintyCNN_LSTMClassifier}. 

\begin{table}
\caption{\small Uncertainty Quantifying in LSTM Classifier \vspace{-1px} }
\label{table:UncertaintyLSTMClassifier}
\begin{center}
\footnotesize
\begin{tabular}{|c||c|c|c|}
  \multicolumn{4}{c}{\textbf{N-BaIoT Dataset}}\\\hline 
  \text{Uncertainty Quantifying}&\text{No Attack} &\text{WBA} &\text{UMD}\\
   \hline
  \text{Deep Ensemble}&98.73&28.63&63.01\\
  \text{Stochastic Weight Averaging}&98.56&28.37&62.69\\\hline
  \multicolumn{4}{c}{\textbf{IoT-23 Dataset}}\\\hline
  \text{Uncertainty Quantifying}&\text{No Attack} &\text{WBA} &\text{UMD}\\
  \hline
  \text{ Method}&&&\\ \hline
  \text{Deep Ensemble}&98.32&31.18&73.74\\
  \text{Stochastic Weight Averaging}&98.29&31.03&73.61\\\hline
\end{tabular}
\end{center}
\end{table}

A comparison of the results of these tables demonstrates that when uncertainty quantification methods are utilized, the accuracy of classification methods is slightly different than when these approaches are not used (in both methods, calculations have been performed 10 times). Furthermore, the consistency of the values in both methodologies demonstrates that the results can be relied on with great certainty.
Table~\ref{table:UncertaintyCNN_LSTMClassifier} presents the results of the uncertainty quantification for the CNN-LSTM classification algorithm, whose results are very similar to those in Table~\ref{table:UncertaintyLSTMClassifier}.
\begin{table}
\caption{\small Uncertainty Quantifying in CNN-LSTM Classifier }
\label{table:UncertaintyCNN_LSTMClassifier}
\begin{center}
\footnotesize
\begin{tabular}{|c||c|c|c|}
  \multicolumn{4}{c}{\textbf{N-BaIoT Dataset}}\\\hline 
  \text{Uncertainty Quantifying}&\text{No Attack} &\text{WBA} &\text{UMD}\\
   \hline
  \text{Deep Ensemble}&99.17&31.06&73.02\\
  \text{Stochastic Weight Averaging}&98.74&30.79&71.82\\\hline
  \multicolumn{4}{c}{\textbf{IoT-23 Dataset}}\\\hline
  \text{Uncertainty Quantifying}&\text{No Attack} &\text{WBA} &\text{UMD}\\
  \hline
  \text{ Method}&&&\\ \hline
  \text{Deep Ensemble}&99.37&33.24&77.09\\
  \text{Stochastic Weight Averaging}&99.08&32.16&76.83\\\hline
\end{tabular}
\end{center}
\end{table}
\section{Conclusions and Future Work}\label{conclusion}
In this paper, we have examined the methods of quantifying uncertainty in deep learning-based botnet detection systems. This paper uses uncertainty in two cases. First, we proposed a defence method against adversarial attacks using uncertainty methods. Second, by applying two uncertainty quantifying methods to the proposed model before and after the attack, their effect on classification was examined. The proposed methods for botnet classification in this paper have an accuracy higher than 98\% (on both datasets), but the proposed attack reduces the accuracy to about 30\%, indicating that the attack was successful. On the other hand, the defence method was designed based on uncertainty and has increased the accuracy to about 70\%, which, although not excellent, is promising.

Due to the growing popularity of distributed systems, some researchers in botnet detection prefer to work on distributed ML techniques like federated learning. Uncertainty in federated learning and the effect of adversarial samples on them are suggested for future work.

\Urlmuskip=0mu plus 1mu\relax
\bibliographystyle{IEEEtran}
\bibliography{main}

\end{document}